\documentclass[aps,twocolumn,showpacs,preprintnumbers,amssymb]{revtex4}

\newcommand{\be}{\begin{eqnarray}}
\newcommand{\ee}{\end{eqnarray}}

\usepackage{dcolumn} 
\def\eea{\end{eqnarray}}
\def\C{\hbox{$\mit I$\kern-.7em$\mit C$}}
\def\N{\hbox{$\mit I$\kern-.3em$\mit N$}}

\usepackage{bm} 

\usepackage{epsfig}

\usepackage{amsmath,amsfonts,amssymb,graphics,graphicx,epsfig,color,times,bbm,epic,eepic}

\begin{document}



\title{Steady state entanglement in open and noisy quantum systems at high temperature}

\author{L. Hartmann$^{1}$, W. D{\"u}r$^{1,2}$ and H.-J. Briegel$^{1,2}$}

\affiliation{$^1$ Institut f{\"u}r Theoretische Physik, Universit{\"a}t Innsbruck,
Technikerstra{\ss}e 25, A-6020 Innsbruck, Austria\\
$^2$ Institut f\"ur Quantenoptik und Quanteninformation der \"Osterreichischen Akademie der Wissenschaften, Innsbruck, Austria.}
\date{\today}

\begin{abstract}
We show that quantum mechanical entanglement can prevail in noisy open quantum systems at high temperature and far from thermodynamical equilibrium, despite the deteriorating effect of decoherence. The system consists of a number $N$ of interacting quantum particles, and it can interact and exchange particles with some environment. The effect of decoherence is counteracted by a simple mechanism, where system particles are randomly reset to some standard initial state, e.g. by replacing them with particles from the environment. We present a master equation that describes this process, which we can solve analytically for small $N$. 
If we vary the interaction strength and the reset against decoherence rate, we find a threshold below which the equilibrium state is classically correlated, and above which there is a parameter region with genuine entanglement. 
\end{abstract}


\maketitle 

While in quantum computation and communication the relevance of entanglement is undisputable and a number of existing protocols use specific entangled states as resources, the situation is not so clear in less controlled, macroscopic systems such as solids or fluids, consisting of a large number of particles, which interact with each other and with the environment.  
Theoretical studies of so-called spin chains and lattices \cite{Sa99,Vi03} show that (bipartite and multipartite) entanglement is present in these systems, at least at very low temperatures when they are close to their energetic ground state \cite{Ar01}. This can be qualitatively understood from the fact that the ground state of most interacting systems is entangled with respect to their interacting constituents. By a continuity argument, then, one expects this to be true also for small temperatures above zero. 
The situation is different, however, for gas--type systems and systems at higher temperatures, or, more generally, systems that are also allowed to exchange particles with the environment and that may be far from thermodynamical equilibrium. Examples of such systems include man-made devices such as the laser \cite{Ha70,Sa74}, but also e.g. bio--molecular processes in cells. 

We study open driven quantum systems from the perspective of quantum information. The main question is whether in interacting systems, where the individual constituents are subjected to decoherence, and where the particle numbers may fluctuate, entanglement can exist in steady--state. Except in situations with very special decoherence mechanisms, or in strongly interacting systems at very low temperature, one expects that decoherence diminishes and inevitably destroys entanglement.
Here, we consider a system of $N$ interacting quantum particles (qubits, for simplicity) and ask whether entanglement can exist under the following conditions. (i) The interaction between the particles is capable of creating entanglement within the system.  (ii) There is an interaction between the particles and an environment, which will lead to decoherence on the state space of the system. (iii) The environment is not allowed to introduce entanglement, i.e. it is supposed to interact only locally with the system particles, leading to individual decoherence. Furthermore, it cannot "import" any entanglement into the system (e.g. by exchanging two system particles with a pair of entangled particles from the environment, nor by some other external control as it is e.g. assumed in quantum error correction). (iv) The system may exchange particles with the environment, in which case the number $N$ may fluctuate around some average value $\bar{N}$. 
An example for a system fulfilling (i-iv) is a spin gas \cite{Ca05}, where particles carrying a spin degree of freedom move freely and interact for a short time upon collision. The particles are subjected to individual decoherence between two collisional events, and also the number of system particles may fluctuate. 
We remark that condition (iii) may not be fulfilled in certain situations, e.g. for strongly interacting systems coupled to a thermal bath. In this situation, interactions between the compound system and the environment may drive transitions to entangled states, allowing for entanglement of the thermal state at low temperatures. However, a local decoherence mechanism as demanded in (iii), which especially takes account of gas--type scenarios where system particles move in a background gas of other molecules, with which they collide, makes it much more difficult to obtain steady--state entanglement.

In this paper, we identify a simple mechanism that satisfies the criteria given above, and which can still lead to entanglement in a steady state not necessarily close to the ground state or thermal state of the interaction Hamiltonian. The mechanism corresponds to replacing randomly (and at a given rate $r$) system particles with particles from the environment in some standard, sufficiently pure, single-particle state. The main purpose of this mechanism is that it locally extracts entropy from the system.
Thereby, existing entanglement between the to-be-replaced particle and the rest of the system is destroyed, but the ``fresh'' particle is ready to become entangled with the system via the interaction Hamiltonian.
This simple mechanism is equivalent to a situation where each particle is measured with the rate $r$ and then reset to a standard state (ignoring the correlations with the other particles). This mechanism, taken alone, certainly can not introduce entanglement into the system, as it acts locally on the individual particles. However, in combination with the interaction between system particles, it is able to create fresh entanglement on an appropriate coarse grained time scale, and thus to counteract the effect of decoherence. The mechanism bears some analogy with driven systems such as an atomic-beam laser, where excited atoms are injected into a laser cavity, providing a resource of energy in form of via stimulated emission of photons, thereby maintaining the state of the laser field. Similar as in laser theory, we find a threshold: If, for a given decoherence and interaction strength, we increase the reset rate $r$, we find a threshold-value below which the steady state of the spins is always classically correlated, while for larger values of $r$ the system shows quantum correlations (i.e. entanglement). 

In the language of reservoir theory \cite{Ha70}, such a system is coupled to two different reservoirs: one at high temperature being responsible for decoherence, and a second at low temperature providing fresh particles with low entropy. The system is thus far away from thermodynamical equilibrium. 
Remarkably, the proposed local(!) reset mechanism allows one to obtain steady state entanglement quite generically, independent of the specific interaction Hamiltonian and of the details of system--environment coupling (e.g. temperature of bath), and also in case of fluctuating particle numbers. This is in contrast to entanglement in thermal states, which is restricted to low temperatures (see e.g. \cite {Ar01}).

In the following, we capture the essential features of such open and noisy systems in a toy model consisting of only two qubits. We give the master equation describing the quantum mechanical interaction, individual noise channels, and the reset mechanism, satisfying conditions (i-iii). We solve the master equation and obtain fully analytic expressions for the entanglement in the system. In a second part, we show that the properties of the toy model carry over to larger systems, even with fluctuating particle numbers.  

{\bf Model.} 
The system consists of $N$ qubits that interact with each other according. We describe this interaction on a coarse--grained timescale by an effective Hamiltonian $H$. This includes gas--type scenarios with short time interactions, as well as continuous coupling between system components.
Additionally, the particles are subject to decoherence described by a noise channel and to a reset channel, modelling the system's interaction with the environment.
We describe the noise channel ${\cal L}_{noise}$ by the Lindblad operator~\cite{DHB}: 
${\cal L}_{noise}\rho=\sum_{i=1}^N -\frac{B}{2}(1-s)[\sigma_+^{(i)}\sigma_-^{(i)}\rho+\rho\sigma_+^{(i)}\sigma_-^{(i)}-2\sigma_-^{(i)}\rho\sigma_+^{(i)}]-\frac{B}{2} s[\sigma_-^{(i)}\sigma_+^{(i)}\rho+\rho\sigma_-^{(i)}\sigma_+^{(i)}-2\sigma_+^{(i)}\rho\sigma_-^{(i)}]-\frac{2C-B}{4}[\rho-\sigma_z^{(i)}\rho\sigma_z^{(i)}],$
where $\sigma_\pm=(\sigma_x \pm i\sigma_y)/2$ and the $\sigma$s are Pauli operators. Parameters $B$ and $C$ give the decay rate of inversion $\langle\frac{1+\sigma_z}{2}\rangle$ and polarization $\langle\sigma_\pm\rangle$, 
and $s=\langle\frac{1+\sigma_z}{2}\rangle_\infty \in [0,1]$ depends on temperature $T$, where $s=1/2$ corresponds to $T=\infty$. This form of the noise channel is based on certain approximations, e.g. Markov approximation. Note, however, that this is not an essential assumption in the present context, but it simplifies the analytic treatment. 
The reset channel corresponds to an additional term specified by Lioville operator ${\cal L}_{reset}$, which we model as follows: 
${\cal L}_{reset}\rho=r\sum_{i=1,2}(|\chi_i\rangle_i\langle\chi_i|\mathrm{tr}_i\rho-\rho)$. 
It means that with some probability $r\delta t$ particle $i$, $i=1...N$, is reset during the time interval $\delta t$ to some specific state $|\chi_i\rangle_i$, while the other qubits are left in the state $\mathrm{tr}_i\rho$. The total master equation is then given by
\begin{equation}\label{GeneralME}
\dot{\rho}=-i[H,\rho]+{\cal L}_{noise}\rho+\sum_{i=1}^N r(|\chi_i\rangle_i\langle\chi_i|\mathrm{tr}_i\rho-\rho).
\end{equation}
We have checked that this equation is of Lindblad form \cite{Li76} and hence it generates a time evolution of $\rho$ described by a completely positive map. For $N=2$, we have analytically solved the equation for Heisenberg and Ising Hamiltonians, $H_H=g \vec\sigma^{(1)} \cdot \vec\sigma^{(2)}$, $H_I=g \sigma_z^{(1)} \sigma_z^{(2)}$, where $g \geq 0$ is the coupling strength.
In the following we will illustrate the essential features for the Ising Hamiltonian $H_I$, and decoherence by a purely dephasing channel. This amounts to setting $B=0$ (then $s$ is arbitrary) and $C=\gamma$, leading to ${\cal L}_{noise}=\gamma/2\sum_{i=1,2}(\sigma_z^{(i)}\rho\sigma_z^{(i)}-\rho)$.
Finally, we consider the case $|\chi_i\rangle_i=|+\rangle_i$, where $\sigma_x|+\rangle=|+\rangle$ is an eigenstate of $\sigma_x$.
The choice of the reset state $|\chi_i\rangle_i$ depends on the Hamiltonian $H$, which should be able to create entanglement from the resulting product state. 
The Ising Hamiltonian could e.g. not create any entanglement if $|\chi_i\rangle_i=|0\rangle$.

The $2$-qubit master equation now takes the following form:
\begin{equation}\label{ME}
\dot{\rho}=-i[H,\rho]+\sum_{i=1,2}\gamma/2(\sigma_z^{(i)}\rho\sigma_z^{(i)}-\rho)+r(|+\rangle_i\langle+|\mathrm{tr}_i\rho-\rho)
\end{equation}
For $r=0$, the steady state of this master equation is diagonal in the product basis $|s_1\rangle|s_2\rangle$, with $s_1,s_2\in\{0,1\}$, and $\sigma_z^{(j)}|s_j\rangle=(-1)^j|s_j\rangle$ ($z$-basis), and there is no entanglement. The dephasing channel, described by the $\gamma$-part, eventually destroys all entanglement that may initially exist or may be created for a short period of time (small compared to $\gamma^{-1}$).

For $r\rightarrow\infty$ we can neglect the Hamiltonian and decoherence parts, and the density matrix will be quasi--permanently projected into the product of two $|+\rangle$-states, and again there will be no entanglement between the qubits. The question is this: Between these two extremes ---a pure product state on the one hand and a classically correlated state on the other hand--- is there a combination of parameters $g$ (Hamiltonian),$\gamma$ (decoherence), and $r$ (reset) such that the steady state will be entangled? Now we show that this is indeed the case.

To solve the master equation~(\ref{ME}), we expand the density operator in the $z$-basis: $\rho (t)=\sum_{s_1^\prime,s_2^\prime,s_1,s_2=0}^1 C_{s_1^\prime,s_2^\prime;s_1,s_2}(t) |s_1^\prime, s_2^\prime\rangle\langle s_1,s_2|$. In this basis, the master equation takes the form of the following $16$ coupled, linear, differential equations for the coefficients: $\dot{C}_{s_1^\prime,s_2^\prime;s_1,s_2}=\{-ig[(-1)^{s_1^\prime+s_2^\prime}-(-1)^{s_1+s_2}]+\gamma/2[(-1)^{s_1^\prime+s_1}+(-1)^{s_2^\prime+s_2}-2]-2r\}C_{s_1^\prime,s_2^\prime;s_1,s_2}+r/2\{C_{0,s_2^\prime;0,s_2}+C_{1,s_2^\prime;1,s_2}+C_{s_1^\prime,0;s_1,0}+C_{s_1^\prime,1;s_1,1}\}$. 
Since the solutions are lengthy we discuss here only the structure. The differential equations for the diagonal elements form a closed set. For any $r>0$ their value is $1/4$ in steady state. For $r=0$ the steady state is not unique since the dephasing channel leaves the diagonal elements untouched, but the entanglement between the two qubits is zero for any initial values. In the following, we will always assume $r>0$. The off-diagonal coefficients that are not on the anti-diagonal are coupled among themselves and are driven by the diagonal elements. Their final values for $t \to \infty$ are $C_{0001}=C_{0010}=C_{0111}^*=C_{1011}^*=\frac{r(-ig+r+\gamma/2)}{4(2g^2+(r+\gamma/2)(r+\gamma))}$. Finally, the anti-diagonal coefficients are not coupled among themselves and are driven by the other off-diagonal elements. 
In steady state, the anti-diagonal is real and contains four times the following entry: $\frac{r^2(r+\gamma/2)}{4(r+\gamma)(2g^2+(r+\gamma/2)(r+\gamma))}$. All other matrix elements are fixed by the Hermiticity of the density matrix. All coefficients approach steady state exponentially fast with their own characteristic exponents determined by the spectrum of the total Liouville super-operator ${\cal L}$ defined by $\dot{\rho}={\cal L}\rho$ with values: $\{0,-r, -2r, -2(r+\gamma), -(3r/2+\gamma\pm 2i\sqrt{g^2-r^2/16})\}$ and  multiplicities $\{1,2,1,4,4+4\}$ respectively.

As a measure for the entanglement between the two qubits we use the negativity~\cite{Vi02}, which is defined with respect to a bipartition $A$-$\bar{A}$ of the set of qubits as ${\cal N}_{A}=(||\rho^{T_{A}}||_1-1)/2$, where $T_A$ means the partial transpose with respect to $A$.
For two qubits, the negativity can assume values between zero (separable state) and $1/2$ (maximally entangled state).

In steady state, we compute from the above expressions for the density matrix the following analytic expression for the negativity in terms of the parameters $g$ (Hamiltonian interaction), $\gamma$ (strength of the dephasing channel), and $r$ (reset rate):

\begin{equation}\label{negativity}
{\cal N}
={\rm max}\{0,-\frac{\gamma(r+\gamma/2)^2+g^2(r+\gamma)-rg(r+\gamma)}{2(r+\gamma)[2g^2+(r+\gamma/2)(r+\gamma)]}\}
\end{equation}
Equation~(\ref{negativity}) contains the full information about the entanglement properties of the two qubits and is one of the main results of this paper. Note that ${\cal N}={\cal N}(\tilde g,\tilde r)$ depends only on two parameters, $\tilde g=g/\gamma$ and $\tilde r=r/\gamma$.

\begin{figure}[ht]
\begin{picture}(230,150)
\put(0,0){\includegraphics[width=0.45\textwidth]{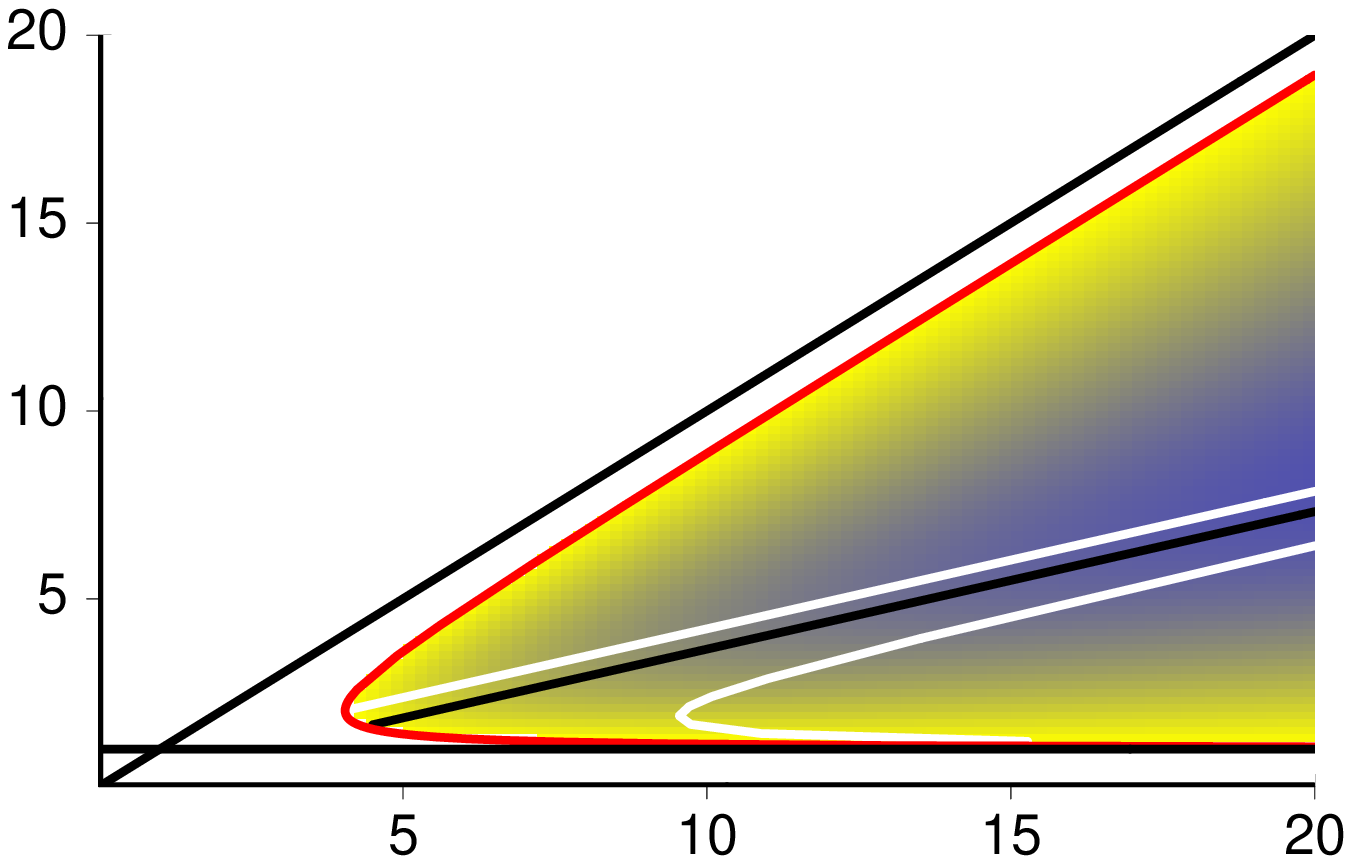}}
\put(25,82){\includegraphics[width=0.2\textwidth]{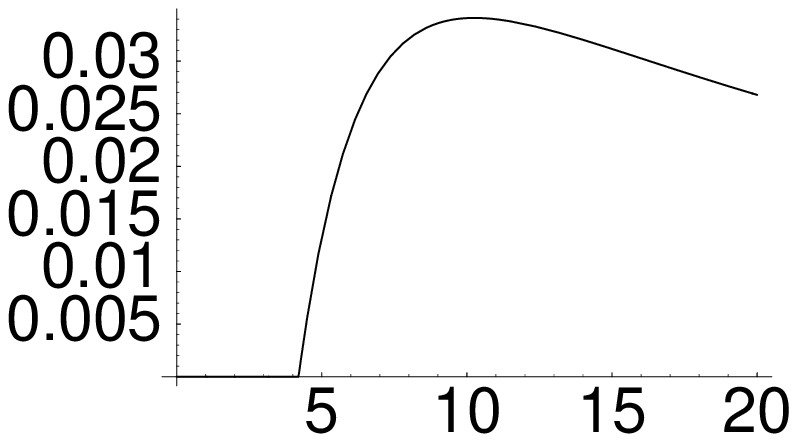}}
\put(82,81){$r/\gamma$}
\put(35,141){${\cal N}$}
\put(115,-8){$r/\gamma$}
\put(-7,72){\Large$\frac{g}{\gamma}$}
\end{picture}
\caption[]{\label{ToyModel} Separable states (white area) and entangled states (colored area) in the $r$-$g$-plane, where $r$ is the rate of our reset process, and $g$ is the coupling strength in the Ising Hamiltonian, and we use $\gamma^{-1}=1$ as unit timescale. The color encodes the amount of entanglement measured by the negativity: the darker the area, the more entanglement is present. For a discussion of the other lines, please see the text. The inlet shows a cut for constant $g=2.5\gamma$.}
\end{figure}

In Fig.~\ref{ToyModel} we see a contour plot of the negativity function. The key feature is the color-coded region in the $r$-$g$-plane with steady state entanglement, where a darker color indicates higher entanglement. Three lines are marked in the entangled region: i) The upper white line is the maximum in $g$-direction (at constant $r$), ii) the lower white line is the maximum in $r$-direction (at constant $g$), and iii) the middle line is the straight $g=r/(1+\sqrt{3})$. This middle line is approached by the upper and lower one asymptotically for large $g$,$r$. The global maximum of the negativity is on this middle line at infinity with a value of approximately $0.092$, about 20\% of the maximal value. The darkest, most entangled area in our plot has negativity $0.068$. The entangled region is bounded by the grey line given by one of the roots of the non-trivial part of equation~(\ref{negativity}). Outside of this region, the state is separable (white area). The entangled region approaches asymptotically the straights $g=\gamma$ and $g=r$ plotted black in Fig.~\ref{ToyModel}. The asymptotic line $g=\gamma$ is independent of $r$ and indicates the border between the weak coupling and the strong coupling regime. 
For weak coupling, $g < \gamma$, decoherence/noise will always triumph over the Hamiltonian part that tries to create entanglement. That is, as a necessary condition, we need to be in the strong coupling limit to observe entanglement. The inlet in Fig.~\ref{ToyModel} shows a cut at $g=2.5\gamma$ through the color-plot. Most notable is the existence of a threshold value for $r/\gamma$ above which entanglement is present in the steady state.

We also simulated the system on a computer, where we use a microscopic decoherence model rather than a master equation description. Decoherence is thereby induced by interactions of system particles with an environment, modelled as a background spin gas~\cite{Ca05}, and we consider interactions of Ising--type, which leads to dephasing noise. Since the spin gas exhibits memory, the decoherence process is non--Markovian and possibly non--local. Also in this case, the results qualitatively agree with the analytic solution of our toy model. We remark that the system itself may also be described as a spin gas, even with fluctuating particle number.

All these results support the following conclusion: Entanglement can persist even in open noisy quantum mechanical systems that are far from thermodynamic equilibrium. It is true that even for $r=0$, i.e. without reset mechanism, one can obtain entangled steady states, e.g. for a decay channel ($s=1,B/2=C=\gamma$) and an interaction Hamiltonian $H=g \sigma_x^{(1)}\sigma_x^{(2)}$. However, this phenomena is restricted to specific Hamiltonians and noise channels, where a steady state of the noise channel with sufficiently low temperature is a necessary condition for steady state entanglement. In particular, entanglement vanishes for high temperatures of the bath, $s \to 1/2$. In contrast, we show in the next section that steady state entanglement appears generically in systems which feature some kind of (appropriate) reset mechanism.  

\begin{figure}[ht]
\begin{picture}(230,85)
\put(115,0){\includegraphics[width=0.225\textwidth]{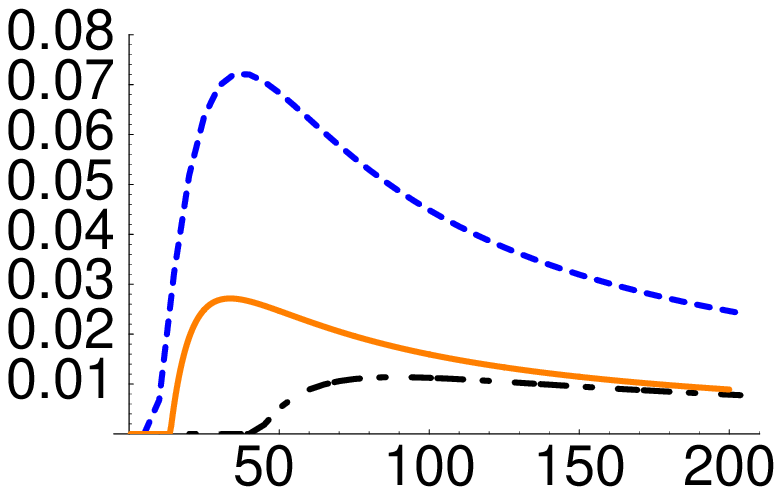}}
\put(-10,-3){\includegraphics[width=0.225\textwidth]{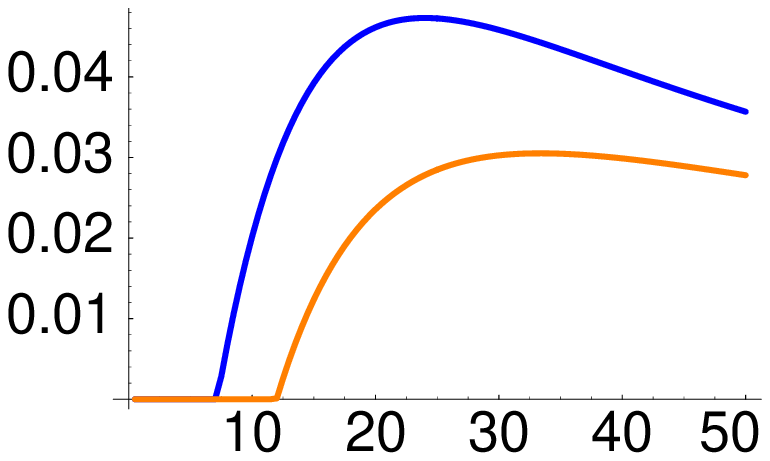}}
\put(50,-5){$r/\gamma$}
\put(-7,75){${\cal N}$}
\put(180,-5){$r/\gamma$}
\put(125,75){$\bar{\cal N}$}
\put(50,75){$(a)$}
\put(180,75){$(b)$}
\end{picture}
\caption[]{\label{morequbits}(a) Negativity for $2$-qubit system with $XYZ$-interaction and magnetic field, $H=g(0.7\sigma_x^{(1)}\sigma_x^{(2)}+0.3\sigma_y^{(1)}\sigma_y^{(2)}+\sigma_z^{(1)}\sigma_z^{(2)}+0.5(\sigma_x^{(1)}+\sigma_x^{(2)}))$, as a function of the reset rate $r$ at $g=10\gamma$. The noise is described by ${\cal L}_{noise}$ of equation~\ref{GeneralME} with $C=B/2$ and $\gamma=C/10$. The upper curve corresponds to zero temperature $(s=0)$, the lower one to infinite temperature $(s=1/2)$ of the bath. Curves for any finite temperature lie in between. (b) Average negativity of (i) $5$-qubits that have all pairwise interacted with each other (dashed), (ii) the reduced density matrix for any two qubits in a $5$-qubit setting (dashed-dotted), and (iii) Poissonian mixture of reduced density matrices corresponding to fluctuating particle number, see text, (solid) as a function of the reset rate $r$ for Ising interaction and dephasing, with $g=5\gamma$.}
\end{figure}

{\bf Beyond toy model: many-particle systems.}
To demonstrate that we have identified a generic feature, we will generalize the system in various directions. 
Although the details will change, the qualitative behavior of the two--qubit model does not depend on the particular choice of the interaction Hamiltonian or the decoherence model. Fig. \ref{morequbits}(a) shows e.g. steady state entanglement for a XYZ-Hamiltonian as function of reset rate $r$, and decoherence described by a noise operator ${\cal L}_{noise}$ with different parameters. The qualitative behavior is similar to the inlet of Fig. \ref{ToyModel}, and we observe steady state entanglement even for infinite temperature of the bath.

Second, the idealized reset mechanism we consider can be replaced by a more realistic imperfect reset mechanism. In this case, fresh particles come in mixed states with sufficiently low entropy rather than in pure states (with entropy 0). Still, the steady state turns out to be entangled.

Third, we have also considered systems consisting of more particles $N$. In this case we are interested in (i) multiparticle entanglement, and (ii) entanglement of reduced states of two qubits. In case of (i), we use the average negativity $\bar{\cal N}$, which is the negativity averaged over all possible bipartitions of the system \cite{Ca05} to measure multiparticle entanglement,
Non--zero average negativity ensures the presence of some form of entanglement in the system. 
To compute this quantity is a formidable task. The system size, and hence the number of differential equations and relevant bipartitions scales exponentially. To simplify computation, we consider a symmetric situation, where all qubits interact pairwise via Ising interactions, and compute $\bar{\cal N}$ of up to seven qubits.
Regarding (ii), we consider pairwise entanglement between two particles in $N$--qubit systems, where the state of the two qubits is given by the reduced density matrix that is obtained by tracing out the remaining particles. Although typically such reduced--state entanglement is rather unstable in multiparticle entangled systems (since entanglement with other system particles acts as additional noise source), we find even in this case a parameter regime with entanglement in steady state. However the region in $r-g$ diagram where one finds entanglement becomes smaller for increasing number of particles $N$. Fig.~\ref{morequbits}(b) shows the plot of the average negativity and negativity of reduced states for systems of $N=5$ qubits. 

Finally, we have considered systems with a fluctuating number of particles $N$. To be precise, we consider systems of $N$ qubits that interact pairwise in a symmetric way, and $N$ fluctuates between $2 \leq N \leq 6$ following a Poissonian distribution in this interval. The state of two qubits, $\rho_{AB}=\sum_{N} p_N \rho_{AB}^{(N)}$, is given by a mixture of reduced density operators $\rho_{AB}^{(N)}$ (obtained from the steady state of the $N$-qubit system) with mixing probabilities $p_N$ given by the probability that the system size is $N$. We find that entanglement between pairs of qubits remains finite (see Fig. \ref{morequbits}(b)).

Generally, for any interacting system subjected to some kind of noise or decoherence (not necessarily local or Markovian), one can expect that an appropriate local reset mechanism leads to entangled steady states, as long as the system is capable of creating entanglement from (slightly mixed) input states at some point in time. For reset rates that are slow enough that entanglement may be built up in the system, but fast enough that decoherence (or other noise processes) will not destroy all the entanglement, one typically expects steady state entanglement on an appropriate coarse grained time scale.

Note that various physical processes are conceivable for the reset mechanisms, including active processes such a measurement or ``optical pumping'' of randomly selected particles in a spin gas, as well as simply replacing a qubit by a fresh one (e.g. spin dependent tunnelling of electrons in charge controlled quantum dots). A process of the latter type may also be conceivable in more natural scenarios such as bio--molecular interactions in a cell. These systems are far away from the thermodynamical equilibrium, and there the fluctuation of particles may, at the same time, act as reset mechanism. 

{\bf Summary:} 
We have shown that entanglement can be present in open noisy quantum systems far from thermodynamic equilibrium. For a two-qubit toy model, we could analytically solve the master equation consisting of a Hamiltonian part, a noise channel, and a proposed reset mechanism, which is responsible for the non-vanishing entanglement in the steady state. We were able to give a closed expression for the entanglement as a function of the parameters of the master equation. We extended the analysis to similar models with other interaction Hamiltonians and decoherence models (including non--Markovian decoherence), systems consisting of more particles, and also systems with fluctuating particle number. In all cases we found that steady state entanglement can prevail. This demonstrates the possibility of stable steady state entanglement in natural systems consisting of many particles which are far away from thermodynamical equilibrium.  

We thank P. Zoller for stimulating discussions. This work was supported by the FWF, the European Union (QUPRODIS, OLAQUI, SCALA), the DFG, and the \"OAW through project APART (W.D.).



\end{document}